\documentclass[sigconf,nonacm]{acmart}
\usepackage{caption}
\usepackage{subcaption}
\usepackage{multirow}
\usepackage{xcolor}
\usepackage{graphicx}
\usepackage{epstopdf}
\usepackage{url}
\usepackage{comment}
\usepackage{booktabs}
\usepackage{microtype}
\usepackage{tabularx}
\AtBeginDocument{%
  \providecommand\BibTeX{{%
  \normalfont
  \kern-0.5em{\scshape i\kern-0.25em b}\kern-0.8em\TeX}
  }}
\acmYear{2026}

\renewcommand\footnotetextcopyrightpermission[1]{}

\graphicspath{{plots/}}
\begin{document}
\raggedbottom

\title{What Makes a Peer? Valuation-Anchored Similarity in Private Markets}


\author{Sebastian Frank}
\email{sebastian.frank@blackrock.com}
\affiliation{%
  \institution{BlackRock, Inc.}
  \city{New York}
  \state{NY}
  \country{USA}
}

\author{Jingrao Lyu}
\email{jingrao.lyu@blackrock.com}
\affiliation{%
  \institution{BlackRock, Inc.}
  \city{Atlanta}
  \state{GA}
  \country{USA}
}

\author{Max Jarmey}
\email{max.jarmey@blackrock.com}
\affiliation{%
  \institution{BlackRock, Inc.}
  \city{London}
  \country{UK}
}

\author{Preetha Saha}
\email{preetha.saha@blackrock.com}
\affiliation{%
  \institution{BlackRock, Inc.}
  \city{New York}
  \state{NY}
  \country{USA}
}

\author{Mingshu Li}
\email{mingshu.li@blackrock.com}
\affiliation{%
  \institution{BlackRock, Inc.}
  \city{Atlanta}
  \state{GA}
  \country{USA}
}

\author{Sweet Kaur}
\email{sweet.kaur@blackrock.com}
\affiliation{%
  \institution{BlackRock, Inc.}
  \city{Gurgaon}
  \state{HR}
  \country{India}
}

\author{Sola Akinola}
\email{sola.akinola@blackrock.com}
\affiliation{%
  \institution{BlackRock, Inc.}
  \city{New York}
  \state{NY}
  \country{USA}
}

\author{Dhagash Mehta}
\email{dhagash.mehta@blackrock.com}
\affiliation{%
  \institution{BlackRock, Inc.}
  \city{New York}
  \state{NY}
  \country{USA}
}

\renewcommand{\shortauthors}{Frank et al.}

\begin{abstract}
As more investors contemplate private markets and contend with limited transparency, sparse disclosures, and infrequent transactions, identifying economically meaningful peer companies for comparison is a fundamental challenge for valuation, due diligence, portfolio construction, and risk management. We propose an ensemble tree-based supervised similarity learning framework that defines company similarity through the lens of market valuation rather than static feature matching or semantic descriptions. Specifically, we train a CatBoost gradient-boosted decision tree model on observed private company valuations and derive a valuation-aware similarity metric from importance-weighted leaf-node co-occurrences across the ensemble. The similarity metric captures shared valuation drivers while accommodating nonlinear relationships, mixed data types, and pervasive missing data common in private markets. Using a global private-market universe of approximately 270,000 companies, including more than 53,000 firms with observed or derivable post-money valuations spanning multiple industries, geographies, and deal stages, we demonstrate that the proposed similarity framework improves upon traditional distance-based and text-embedding-based approaches in downstream k-nearest-neighbor valuation tasks in the evaluated industry groups, while retaining case-based explainability.

\end{abstract}

\maketitle

\section{Introduction}

Private markets have become one of the fastest-growing segments of global financial markets spanning private equity (PE), venture capital (VC), private credit, real estate, and other asset classes. VCs, a subset focused on new or growing companies, invest in early-stage private companies with high growth expectations and correspondingly elevated risk profiles. Despite experiencing a $5\%$ decrease in buyout volume in 2025, the PE market recorded the largest deals in its history, a 40\% increase in PE-backed exits (100\% for IPOs), and a 17\% increase in overall deal value \cite{edlich2026privateequity}. In spite of their growing importance, private companies remain inherently challenging to analyze at scale due to sparse corporate disclosures, irregular reporting intervals, diverse business models, and infrequent transaction data. 

As private markets expand, the availability of consistent, explainable analytics that standardize how practitioners benchmark and compare private assets has become critical for workflows spanning investment research, due diligence, and risk-oriented quality control. The identification of economically meaningful comparables determining which companies constitute true peer assets lies at the core of these allocation decisions \cite{damodaran2011damodaran}. Investors implicitly rely on peer groups to contextualize valuations, assess growth trajectories, evaluate competitive positioning and perform due diligence. Traditionally, these peer groups are constructed using coarse filters such as industry classification, verticals, geography, deal stage or product descriptions. While intuitive, such rule-based approaches struggle to balance relevance and coverage where broad heuristics admit weakly related firms and narrow filters risk excluding informative peers altogether.

Recent advances in supervised similarity learning have motivated a reassessment of how firm similarity is defined and learned \cite{Jeyapaulraj2022,geertsema2023relative,Saha2024,rosaler2025supervised}. Rather than treating similarity as a static distance in feature space, supervised approaches learn similarity with respect to an explicit objective. In this setting, firms are considered similar because they jointly explain an outcome of interest.

Compared with unsupervised distance metrics, supervised similarity learning learns feature relevance directly from the target variable, reducing reliance on manual weighting and domain expertise. The resulting similarity structure is task-specific and supports applications such as peer identification, localized valuation, anomaly detection, and data quality assessment. For private markets, where data sparsity and dispersion is pervasive, this provides a principled and scalable framework for company comparison and peer discovery anchored on a predetermined objective or the target variable.

A valuation-conditioned similarity metric has direct relevance for private-market workflows beyond valuation modeling. In sourcing, investors can scan for companies that resemble high-performing portfolio firms or recently transacted peers, surfacing opportunities outside conventional sector boundaries. In benchmarking and diligence, analysts can compare a target's valuation multiples and growth profile against its learned nearest neighbors, providing a market-grounded reference that complements manual comparable-company selection. In portfolio construction and risk management, the similarity structure can help identify concentration risk among holdings that appear diversified by industry but share similar valuation drivers, and can inform secondaries and fund-level analytics by grouping assets according to shared pricing dynamics.

\subsection{Related Work}
Private-company valuation commonly relies on discounted cash flow analysis, transaction multiples, and comparable-company methods, all of which depend critically on selecting economically relevant peers \cite{damodaran2011damodaran}. Recent work has used ML to make this selection data-driven: Ref.~\cite{geertsema2023relative} jointly addressed relative valuation and peer-firm selection by learning nonlinear valuation functions and expressing predicted valuation multiples as weighted averages of peer-firm multiples. Their results demonstrate that ML can recover economically meaningful peer weights and improve out-of-sample valuation accuracy. In private-company settings, Ref.~\cite{zhang2023application} similarly showed that ML models can improve entrepreneurial-firm valuation when conventional financial information is limited. Our work builds on this literature but differs in objective: rather than using peer weights primarily to produce a valuation estimate, we explicitly construct a reusable pairwise similarity structure anchored to observed private-company valuations.

A complementary literature represents inter-company relationships directly. CompanyKG \cite{cao2024companykg} models firms and their heterogeneous relationships as a large-scale graph and evaluates learned representations on company-similarity tasks. Ref.~\cite{vamvourellis2024company} used pretrained and fine-tuned language models to derive company embeddings from business descriptions in SEC filings, showing that the resulting representations recover industry structure and identify firms with similar financial characteristics and return correlations.
 More generally, metric learning seeks to infer task-specific notions of proximity from supervised information rather than impose a fixed geometric distance \cite{bellet2014survey}. This distinction is particularly relevant to private markets, where industry taxonomies are non-unique, firm descriptions are incomplete, and the appropriate definition of a peer depends on the downstream economic objective.

Tree ensembles provide a natural mechanism for learning such task-conditioned similarities. Breiman\cite{Breiman2001Random} introduced random-forest  proximity\footnote{\url{https://www.stat.berkeley.edu/~breiman/RandomForests/cc_home.htm}}, in which observations are considered similar when they repeatedly occupy the same terminal nodes. Subsequent work has formalized tree-induced proximity as a representation of the supervised geometry learned by an ensemble \cite{rhodes2023geometry}. In finance, tree-based proximities have been applied to corporate-bond similarity \cite{jeyapaulraj2022supervised} and relative valuation in illiquid municipal-bond markets \cite{Saha2024}. Our work extends this line of research to private-company equity and venture-capital settings by learning a valuation-conditioned similarity structure from diverse company and transaction attributes. Unlike methods that use similarity only for peer filtering or as an implicit component of prediction, the proposed framework explicitly constructs an tree-importance-weighted pairwise similarity matrix whose geometry is anchored to observed post-money valuations.

\subsection{Key Contributions}
This paper introduces an ML framework for identifying valuation-relevant peer companies in private markets. Although the framework is trained using post-money valuation as the supervisory target, its primary objective is not standalone valuation prediction but the extraction of a valuation-conditioned similarity structure among private companies. We formalize company similarity as a supervised learning problem, where similarity is learned from observed market valuations rather than predefined distance functions, industry classifications, or textual descriptions. By anchoring similarity to valuation outcomes, the proposed approach generates peer relationships that are directly aligned with private market pricing dynamics.

The key contributions of this work are threefold. First, we develop a supervised similarity framework based on CatBoost models trained on post-money valuations and derive a tree-importance-weighted leaf-node co-occurrence metric that produces a symmetric pairwise similarity matrix across companies. Second, we demonstrate that the learned metric improves upon traditional unsupervised and semantic similarity approaches, including Euclidean, Gower, and embedding-based distances described in Table \ref{tab:distance_metrics_knn}, in downstream k-nearest-neighbor valuation tasks within the evaluated industry groups. These results are consistent with the broader distance metric learning literature, which shows that task-specific similarity functions often outperform generic geometric distance measures \cite{xing2002distance,weinberger2009distance}. Finally, we provide interpretable explanations of both valuation predictions and peer relationships using SHAP \cite{Lundberg2020From}, highlighting the valuation drivers that shape the induced similarity structure.

\begin{table}[htbp]
\centering
\caption{Distance metrics compared in k-NN evaluation}
\label{tab:distance_metrics_knn}
\begin{tabularx}{\columnwidth}{lX}
\toprule
\textbf{Metric} & \textbf{Description} \\
\midrule
Cosine & $1 - \frac{x \cdot y}{\|x\| \cdot \|y\|}$ over MPNet dense embeddings \\
\addlinespace
Euclidean & $\sqrt{\sum (x_j - y_j)^2}$ over standardized + LOO encoded features \\
\addlinespace
Gower & $(1/p) \sum_j d_j(x_j, y_j)$ with mixed-type normalization \\
\addlinespace
Learned (Ours) & $1 - \sum_t w_t \cdot \mathbb{I}[Z_{1,t} = Z_{2,t}]$ (Proposed framework) \\
\bottomrule
\end{tabularx}
\end{table}

\section{Private Company Data}
Our dataset comprises approximately 270,000 private companies globally, spanning more than 50 industry groups, hundreds of product and service categories, and multiple geographic dimensions including region, country, state, and city. Deal history in the dataset extends from 1977 to 2025, with the median deal having an age of five years as of calibration (2025), and 75\% of deals having taken place since 2015. Table \ref{feature-table} shows the key variables available for our experiments. The dataset contains company-level characteristics, deal information from the most recent funding round, observed transaction-based valuations, and the latest available financial data.

\subsection{Feature description}

Post-money valuation, or the post-deal equity valuation of a firm, is available following a recent funding round. For some companies, we see a post-money valuation directly, while for others, we compute this amount based on enterprise value, deal size divided by the acquisition percentage, pre-money valuation (equity valuation prior to the deal) plus deal size, or pre-money valuation plus the total offering amount. This enables us to recover approximately 53,000 post-money valuations (the most recent deal for each firm in our universe). 
The final feature set contains 27 variables, of which 23 are categorical, reflecting the inherently taxonomic nature of private market data. Key attributes such as deal stage, deal type, internal sector classification, geography, customer profile, and product offerings are naturally categorical and often exhibit high cardinality. For example, deal type contains approximately 80 distinct transaction categories spanning venture capital and private equity markets, while internal sector and industry classifications encompass hundreds of fine-grained categories. In addition, several variables contain multi-label assignments, further increasing the complexity of the feature space.

A defining characteristic of private-market data is the prevalence of missing information, reflecting limited disclosure requirements for private companies in jurisdictions such as the United States. Of the approximately 270,000 companies in the original universe, roughly 53,000 have an observed or derivable post-money valuation and at least one PE or VC transaction, forming the calibration universe for this study. Financial variables such as revenue (top-line sales), EBITDA (earnings before interest, taxes, depreciation, and amortization), and net income (income after all expenses) exhibit the highest levels of missing values (75--80\%), whereas categorical attributes including country, deal stage, and customer type are largely complete. Feature cardinality is highest among geographic and sector-related variables and lowest among strategy, status, and offering-type fields.

To construct the final feature set, we applied a two-stage selection procedure. First, categorical features with fewer than 50\% non-missing observations were excluded. Second, features exhibiting zero importance in a preliminary CatBoost model were removed. This process reduced the risk of learning unstable decision boundaries from missing feature values. Summary statistics for the principal features are provided in Table~\ref{tab:features}.


\begin{table*}[t]
\centering
\caption{Feature categories and representative variables (Metrics from Calibration)}
\label{feature-table}
\resizebox{\textwidth}{!}{%
\begin{tabular}{lll}
\toprule
\textbf{Category} & \textbf{Type} & \textbf{Key Features (\% Missing Values for Numerical Features, Cardinality for Categorical Features)}\\
\midrule
Financial & Numerical & Revenue (78\%), EBITDA (80\%), Net Income (78\%), Avg Time Since Last Financials (75\%) \\
\addlinespace
Deal-level & Categorical & Deal Type (83), Strategy (2), Target Status (2), Offering Type (3) \\
\addlinespace
Geographic & Categorical & Country (130), Region and City (4,816) \\
\addlinespace
Sector/Product Industry & Categorical & Internal Industry Group (720), Internal Sector Categorization (229), Product/Service Offering (305) \\
\addlinespace
Customer & Categorical & Customer Type (4), Customer End User (7), Customer Use (244)\\
\bottomrule
\end{tabular}}
\label{tab:features}
\end{table*}


\section{Valuation-based similarity learning framework}\label{sec:proposed_method}
\subsection{Valuation as a Supervised Anchor}
A central challenge in private markets is that there is no universally accepted definition of what constitutes a comparable company \cite{mehta2025clustering}. Traditional approaches rely on industry classifications, geography, funding stage, or business descriptions, yet private companies frequently span multiple sectors, products, and customer segments simultaneously. As a result, peer identification is often subjective and highly dependent on practitioner judgment.

We use observed private market valuations as the supervisory anchor for learning company similarity. Unlike predefined taxonomies, valuations reflect investors' assessment of a firm's growth prospects, competitive position, financial performance, and risk profile at the time of a transaction \cite{damodaran2011damodaran}. As such, valuation provides a market-based summary of company characteristics and future expectations.

Conditioning similarity on valuation aligns peer identification with the primary objective of comparable analysis. Under this framework, companies are considered peers because they exhibit similar valuation dynamics given their underlying attributes. Importantly, the framework does not simply group firms with similar valuation levels. Companies with identical valuations may be deemed dissimilar if their valuations are driven by different factors, while firms with different valuations may be identified as close peers if they share similar valuation drivers. The resulting similarity metric therefore captures common valuation mechanisms rather than merely matching firms by size or market value.

\subsection{Tree-Based Similarity Learning via Gradient Boosted Decision Trees}

Gradient boosted decision trees (GBDTs) are well suited to private market data because they naturally accommodate mixed data types, missing values, and complex nonlinear interactions \cite{Jeyapaulraj2022,Saha2024}. When trained to predict company valuations based on multiple input features, GBDTs partition firms into locally homogeneous regions defined by valuation-relevant feature interactions, including geography, sector, deal stage, and financial attributes. This learned structure provides a natural foundation for supervised similarity learning: firms are considered similar if they are repeatedly grouped together by the valuation model.

To construct the similarity metric based on the trained GBDT, we leverage leaf-node co-occurrences across the ensemble. Because GBDTs are trained sequentially to minimize prediction error \cite{Friedman2001Greedy}, different trees contribute unequally to the final model. We therefore weight each tree according to its incremental reduction in training loss, assigning greater influence to trees that contribute more to predictive performance. Pairwise similarity is then computed as the weighted frequency with which two companies are assigned to the same terminal node across the ensemble.

The resulting similarity matrix captures shared valuation drivers rather than simple feature proximity. By conditioning similarity on the valuation objective, the framework learns a localized, nonlinear distance metric that identifies companies exhibiting similar valuation behavior.

\subsection{CatBoost Implementation}

We implement the valuation model using Categorical Boosting (CatBoost)\footnote{\url{https://catboost.ai/}}, a gradient-boosted decision tree algorithm designed for structured datasets containing high-cardinality categorical variables and missing information \cite{prokhorenkova2018catboost}. This setting closely matches private market data, where economically important attributes such as deal type, sector classification, product offering, customer segment, and geography are predominantly categorical and often contain hundreds or thousands of distinct values.

Unlike conventional boosting pipelines that require one-hot or externally computed target encodings, CatBoost processes categorical variables natively through ordered target statistics. These statistics are constructed using only observations that precede the current instance under a random permutation, thereby reducing target leakage and the prediction shift associated with standard target encoding. CatBoost further employs ordered boosting, in which residual estimates are generated without using the target value of the observation being predicted. Together, these mechanisms provide a leakage-resistant treatment of high-cardinality categorical information while avoiding the dimensional expansion induced by one-hot encoding.

CatBoost also handles missing numerical values directly through learned split directions, allowing missingness to contribute predictive information without explicit imputation. The model is built from symmetric, or oblivious, decision trees, in which the same split condition is applied at each depth. This structure enables efficient inference, regularization, and straightforward extraction of terminal-node assignments across the ensemble. The latter property is central to our framework, since company similarity is derived from tree-importance-weighted leaf-node co-occur\-rences. CatBoost therefore serves not only as the valuation estimator, but also as the learned partitioning mechanism from which the valuation-con\-ditioned similarity structure is constructed.

\subsection{Regression}
Let \(z_i=\log(y_i)\) denote the log-transformed post-money valuation. The model is trained to minimize the root mean squared error in log space:
\begin{equation}
\mathcal{L}
=
\sqrt{
\frac{1}{N}
\sum_{i=1}^{N}
\left(\hat{z}_i-z_i\right)^2
}.
\label{eq:reg_loss}
\end{equation}

Here, \(z_i\) and \(\hat{z}_i\) denote the observed and predicted log valuations, respectively. The predictive ensemble function \(\hat{z}(x)\) is constructed sequentially as an additive sum of \(T\) decision trees scaled by the learning rate \(\eta\):
\begin{equation}
\hat{z}(x) = \sum_{t=1}^{T}\eta\,h_t(x),
\label{eq:ensemble}
\end{equation}
where \(h_t(x)\) denotes the output of tree \(t\). Predictions are transformed back to valuation levels as
\(\hat{y}(x)=\exp(\hat{z}(x))\).

\subsubsection{Tree Importance Weighting}
Not all trees contribute equally to predictive accuracy. Earlier trees typically capture dominant predictive structure, while later trees refine residual errors. We assign importance weights based on the incremental reduction in loss.
Let $TL_t$ denote the training loss after the first $t$ trees, where $TL_0$ represents the loss of the initial null model. The step-size predictive contribution $S_t$ of an individual tree $t$ is calculated via the absolute difference in sequential training loss:
\begin{equation}
S_t = |TL_t - TL_{t-1}|
\end{equation}

To establish the proportional influence of each weak learner on the final similarity space, the normalized tree weight $w_t$ is computed as:
\begin{equation}
w_t = \frac{S_t}{\sum_{j=1}^T S_j}
\end{equation}
This ensures that trees contributing the largest improvements to the loss have proportionally greater influence on the similarity measure.

\subsubsection{Leaf-Node Co-occurrence Similarity and Supervised Dissimilarity}
For companies $X_1$ and $X_2$, let $Z_{1,t}$ and $Z_{2,t}$ denote leaf nodes to which they fall into while traversing an assigned tree \textit{t} in an ensemble. The tree-weighted leaf-node co-occurrence similarity function $S(X_1, X_2)$ is formulated as:

\begin{equation}
S(X_1, X_2) = \sum_{t=1}^T w_t \cdot \mathbb{I}[Z_{1,t} = Z_{2,t}]
\end{equation}
where $\mathbb{I}[\cdot]$ is the indicator function equal to 1 when the two companies occupy the same terminal node and 0 otherwise.

The final bounded supervised dissimilarity measure $D(X_1, X_2)$ is derived directly from the complement of the similarity matrix score:
\begin{equation}
D(X_1, X_2) = 1 - S(X_1, X_2)
\end{equation}
This operation maps the pairwise relationship into a symmetric matrix where $D \in [0, 1]$.

While we refer to the proposed measure as a distance, it is best viewed as a supervised dissimilarity function that captures valuation-conditioned proximity; it is not guaranteed to satisfy all metric-space axioms such as the triangle inequality.

\section{Experimental Setup}
For training the CatBoost algorithm, we use the cross-sectional company and transaction attributes as input variables shown in Table \ref{feature-table}. We use post-money valuation associated with the most recent observed transaction as the target variable. Post-money valuation is the implied equity value of a company immediately after its most recent financing round, calculated by multiplying the total shares outstanding (common stock, preferred stock, and option pool) by the conversion price per share of that round, and may be sourced directly from press releases, computed from issued shares in regulatory filings, or estimated from authorized shares with a discount factor.

\subsection{Feature Engineering and Preprocessing}

\subsubsection{Target Variable}
Given the highly right-skewed nature of private company valuations, the target variable is transformed into log space prior to model calibration and winsorized at the 1st and 99th percentiles to mitigate the influence of extreme outliers.

The target valuation is constructed from multiple transaction-related fields. When available, post-money valuation is used directly. Otherwise, it is derived from enterprise value, acquisition transaction terms, pre-money valuation plus deal size, or pre-money valuation plus total offering amount. The model is trained to minimize the root mean squared error (RMSE) of log valuations (Eq. \ref{eq:reg_loss}), which stabilizes variance and improves predictive performance across the wide range of company sizes observed in private markets.




\subsubsection{Input Variables}

Input variables (Table \ref{feature-table}) undergo minimal preprocessing to preserve their economic interpretation while ensuring robustness. Multi-label deal types are concatenated into a single categorical feature prior to training. Numerical variables are retained in their original scale, while categorical features are passed directly to CatBoost, which natively handles high-cardinality categorical variables through ordered target statistics. Missing values are not explicitly imputed and are instead incorporated into the tree-splitting process within CatBoost as potentially informative signals.
In addition, we construct an \textit{Average Time Since Last Financials} feature to capture the staleness of reported financial information. For each financial metric (revenue, EBITDA, and net income), we measure the time elapsed since they were last reported, then take the average of those time horizons. Because the dataset represents a snapshot in time as of September 2025, this variable enables the model to distinguish between structurally missing financials and observations whose financial data may simply be outdated.


\subsection{Modeling}
\subsubsection{Stratified Sampling}

The dataset is divided into training and test sets using an 80:20 split stratified by internal subindustry classification to preserve category representation across both samples. The same stratification scheme is applied within cross-validation folds used for hyperparameter optimization. All preprocessing decisions, feature selection, model estimation, and tree-importance weights are derived exclusively from the training data. After fitting, the trained CatBoost ensemble is held fixed and used to compute similarities between test companies and the training reference set. Test-set valuations are used only for final performance evaluation and do not influence model fitting, neighbor selection, similarity weights, scaling-factor estimation, or conformal calibration.
 

\subsubsection{Sample Weighting}
Given that the historical transactions in the calibration dataset span nearly fifty years (median: five years), we apply recency-based weighting to the training samples to emphasize more recent transactions. Specifically, sample weights increase with transaction recency, reflecting structural changes in private markets and the time-varying nature of valuation determinants. Four observations with missing transaction dates were assigned the minimum recency, putting their weights in line with the median sample weight.

\subsubsection{Hyperparameter Optimization}
We optimize hyperparameters using Optuna v4.7.0 \cite{akiba2019optuna} with tree-structured Parzen Estimator (TPE) sampling over 50 trials using five-fold cross-validation on the training set. The optimal configuration and search ranges are shown in Table \ref{tab:hyperparams}.

\begin{table}[htbp]
\centering
\caption{Optimized CatBoost Hyperparameters via Optuna}
\label{tab:hyperparams}
\resizebox{\columnwidth}{!}{%
\begin{tabular}{llll}
\toprule
\textbf{Hyperparameter} & \textbf{Best Value} & \textbf{Search Range} & \textbf{Step Size} \\
\midrule
Tree Depth & 13 & $[7, 14]$ & 1 \\
\addlinespace
Number of Estimators & 1,859 & $[500, 2000]$ &  1\\
\addlinespace
Max Leaves per Tree & 108 & $[31, 127]$ &  1\\
\addlinespace
Min Data in Leaf & 242 & $[128, 512]$ &  1\\
\addlinespace
Bagging Temperature & 6.96 & $[1, 8]$ &  Cont.\\
\addlinespace
Random Strength & 7.71 & $[2, 10]$ &  Cont.\\
\addlinespace
Learning Rate ($\eta$) & 0.023 & $[0.015, 0.06]$ &  Cont. (log)\\
\addlinespace
Max Bin & 164 & $[128, 255]$ &  1\\
\addlinespace
L2 Leaf Regularization & 24.28 & $[10, 200]$ &  Cont. (log)\\
\bottomrule
\end{tabular}}
\end{table}

\subsection{Model Outputs}
\subsubsection{Valuation Regression and Benchmark Models}
The valuation forecast produces a single point estimate in log-space, which can then be exponentiated and scaled to a linear value. The model accuracy is measured based on mean absolute error (MAE), RMSE, $R^2$, mean absolute percentage error (MAPE) and median absolute percentage error (MdAPE). We also used Ordinary Least Squares (OLS) as a baseline before selecting the CatBoost model for valuation prediction. Leave one out encoding was used on the categorical inputs, while simple mean imputation was performed for missing numeric data.

\subsubsection{Conformal Prediction}
To quantify predictive uncertainty, we use conformal prediction, which constructs distribution-free prediction intervals from calibration residuals \cite{shafer2008tutorial,li2025similarity}. Residuals are normalized and grouped by deal type to account for heteroskedasticity across funding rounds. The resulting intervals provide calibrated uncertainty bounds around each valuation estimate.

\subsubsection{Similarity Matrix}
Beyond valuation prediction, the trained model is used to construct a pairwise similarity matrix across companies. Similarity is computed as the importance-weighted frequency with which two firms are assigned to the same terminal node across the ensemble. Trees that contribute more to predictive performance receive greater weight, ensuring that the resulting similarity measure reflects the most influential valuation drivers learned by the model.

\section{Results}
\subsection{Out-of-Sample Model Performance}
The CatBoost model outperforms a multivariate linear regression. We evaluate model performance using standard regression metrics.
\begin{table}[H]
    \centering
    \caption{Regression metrics and comparison with benchmark model.}
    \label{tab:oos_metrics}
    \begin{tabular}{lrrr}
        \toprule
        \textbf{Metric} & \textbf{OLS} & \textbf{CatBoost} & \textbf{\% Relative Improvement} \\
        \midrule
        MAE             & 1.19  & 1.08  & 8\% \\
        RMSE            & 1.56  & 1.44  & 8\% \\
        $R^2$           & 0.36  & 0.46  & 28\% \\
        MAPE (\%)       & 0.07  & 0.06   & 14\% \\
        MdAPE (\%)      & 0.05  & 0.05   & -\% \\
        \bottomrule
    \end{tabular}
\end{table}

To assess model performance across the target distribution, we partition the test set into quintiles based on the ground-truth log valuation. We further disaggregate performance by categorical features to identify segments where the model exhibits systematic bias. 
Prediction errors are concentrated in the lowest and highest valuation deciles (the smallest and largest sized firms). Out-of-sample RMSE rises from 0.94 in Q8 to 2.35 in Q1 and 2.05 in Q10, with an overall RMSE of 1.44, indicating that errors are largest for the smallest and largest firms and lowest in the middle of the valuation distribution.

Performance also varies across geographies, industries, and transaction types. The ten most represented countries account for approximately 90\% of the test dataset. Higher prediction errors are observed in cybersecurity and selected software segments, reflecting the greater dispersion and uncertainty of valuations within these sectors. Across transaction types, the most prevalent categories represent approximately 83\% of the out-of-sample dataset. Prediction errors are highest for later-stage private equity transactions and deals classified as unspecified, whereas early-stage VC financings tend to exhibit lower errors and more consistent valuation patterns.

\subsection{Model Scaling}
Because the model is estimated in log-valuation space, directly exponentiating predictions introduces downward retransformation bias in linear space. We correct for this using multiplicative calibration factors estimated on the training data as the ratio of aggregate observed valuations to aggregate exponentiated predictions. To account for differences in valuation distributions across transaction categories, separate factors are applied by deal type: 1.74 for Series transactions, 5.06 for Unspecified transactions, and 2.19 for all other transactions. These adjustments improve calibration of predicted valuations in level terms.

\subsection{Conformal Prediction}
\begin{figure}[H]
    \centering
    \includegraphics[width=\columnwidth]{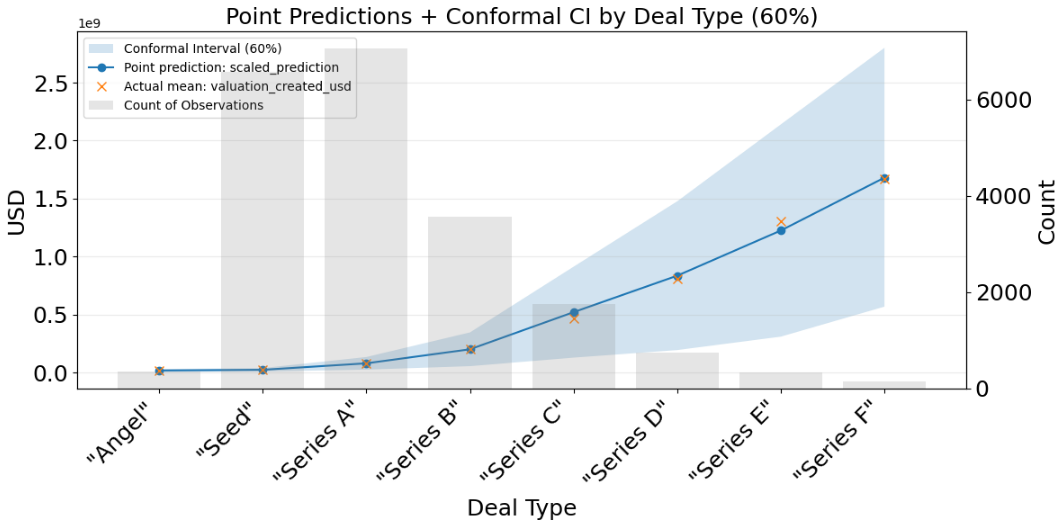}
    \caption{Conformal Prediction Intervals by Deal Stage}
    \label{fig:conformal_pred}
\end{figure}

Figure \ref{fig:conformal_pred} illustrates conformal prediction intervals across funding stages. The intervals capture a substantial proportion of observed valuations while providing calibrated uncertainty estimates around each prediction. Interval widths generally increase in later funding stages, consistent with the greater dispersion and uncertainty associated with valuations of mature private companies. As expected, higher confidence levels produce wider intervals, reflecting the trade-off between coverage and interval efficiency.
 
Conformal metrics for three different confidence levels are displayed below. Coverage percentage represents the proportion of observed valuations captured by the conformal interval, while interval mean ratio refers to the average interval size relative to the mean prediction. 
\begin{table}[h]
\centering
\caption{Prediction interval statistics}
\label{tab:prediction_intervals}
\resizebox{\columnwidth}{!}{%
\begin{tabular}{lrrr}
\toprule
\textbf{Count} & \textbf{Confidence Level} & \textbf{Coverage \%} & \textbf{Interval Mean Ratio} \\
\midrule
53.8k & 95\% & 93\%  & 3.35  \\
53.8k & 75\% & 74\%  & 1.80 \\
53.8k & 60\% & 60\%  & 1.60  \\
\bottomrule
\end{tabular}}
\end{table}

\section{Model interpretation and similarity evaluation}

We use SHapley Additive exPlanations (SHAP)\footnote{\url{https://shap.readthedocs.io/}} to interpret the ML model \cite{lundberg2017shap,yang2021fast}. Global feature importance is measured by the mean absolute SHAP value across observations, capturing each feature's average contribution to predicted valuation.

\begin{figure}[H]
    \centering
    \includegraphics[width=\columnwidth]{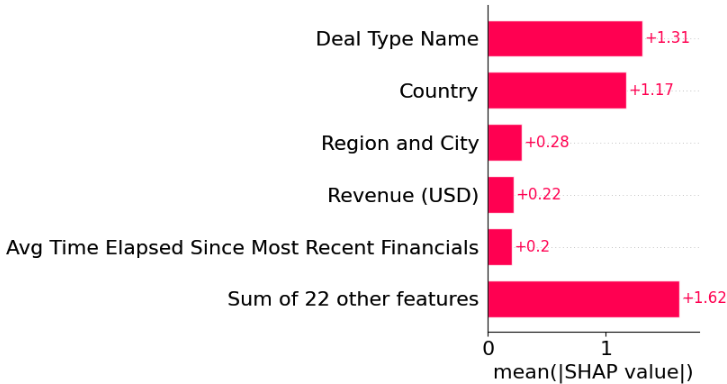}
    \caption{Bar Plot of Absolute Mean SHAP values}
    \label{fig:shap}
\end{figure}

Figure~\ref{fig:shap} shows that deal type is the dominant valuation driver, with a mean absolute SHAP contribution of 1.31, followed by country (1.17), region and city (0.28), revenue (0.22), and average time elapsed since most recent financials (0.20). The remaining 22 features contribute a combined 1.62. These top five features therefore play the largest role in shaping both valuation predictions and the induced similarity structure.

\subsection{Neighborhood Analysis}
The objective of the similarity evaluation is to determine whether the learned metric produces economically coherent peer groups and captures meaningful relationships between private companies. We assess this both qualitatively, through neighborhood consistency, and quantitatively, through comparisons with established distance and embedding based baselines. A successful similarity framework should identify peers that are more similar across economically relevant characteristics than those obtained from conventional proximity measures.

To evaluate neighborhood consistency, we identify the 20 nearest neighbors of each company under the learned similarity metric. For numerical attributes, we compute the average absolute deviation between a company and its neighbors, while for categorical attributes we calculate the proportion of mismatched categories. These measures are aggregated across all companies within an industry group and compared against baseline similarity methods. Lower numerical deviations and categorical mismatch rates indicate that the learned similarity metric produces more coherent peer groups with respect to economically meaningful company characteristics.

\begin{figure*}[h]
    \centering
    \includegraphics[width=\textwidth]{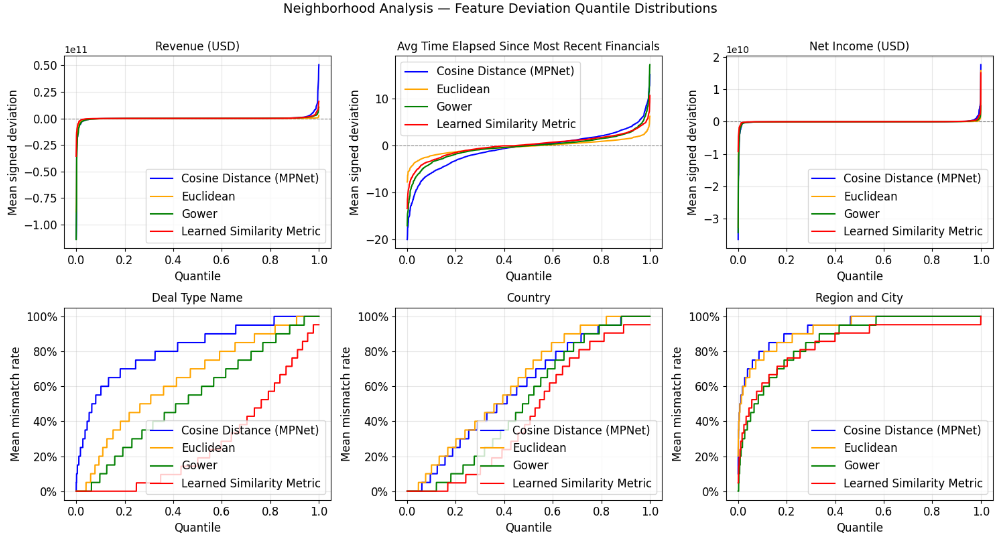}
    \caption{Financial Services (3.2k firms)}
    \label{fig:similarity_benchmark_knn}
\end{figure*}

The comparison with other distance metrics (Figure~\ref{fig:similarity_benchmark_knn}) shows that for the top three numeric features as measured by SHAP (revenue, average time elapsed since last financials, and net income), the learned metric consistently shows a lower deviation than cosine distance (between embeddings derived from internal company descriptions) and mixed performance against Gower and Euclidean distances on Financial Services companies. For the top three categorical features (deal type, country, and region and city), the learned metric shows a lower mismatch rate than other metrics.

\subsection{k-NN Benchmarking}
To evaluate the learned similarity metric further, we compare its ability to recover company valuations against several benchmark distance measures, including Euclidean, Gower, and embedding-based similarity. For each metric, we construct a weighted k-nearest-neighbor (k-NN) estimator and assess valuation prediction accuracy across multiple neighborhood sizes. Superior performance indicates that the corresponding similarity measure identifies firms with more comparable valuation characteristics.

While the proposed similarity metric is learned from valuation data, this evaluation does not reuse CatBoost valuation predictions: the CatBoost model is used only to construct the similarity structure, whereas valuation estimates are generated independently through a non-parametric k-NN procedure operating on the resulting neighborhood graph. The experiment therefore evaluates whether the learned similarity metric produces neighborhoods that are more valuation-consistent than those obtained from conventional distance measures. If the similarity structure is economically meaningful, firms identified as neighbors should exhibit similar valuations, leading to improved k-NN performance. The results demonstrate that the learned metric outperforms traditional distance-based and semantic baselines in the evaluated settings, suggesting that it captures valuation-relevant relationships more effectively than generic notions of proximity in those settings.

\begin{figure*}[h]
    \centering
    \includegraphics[width=\textwidth]{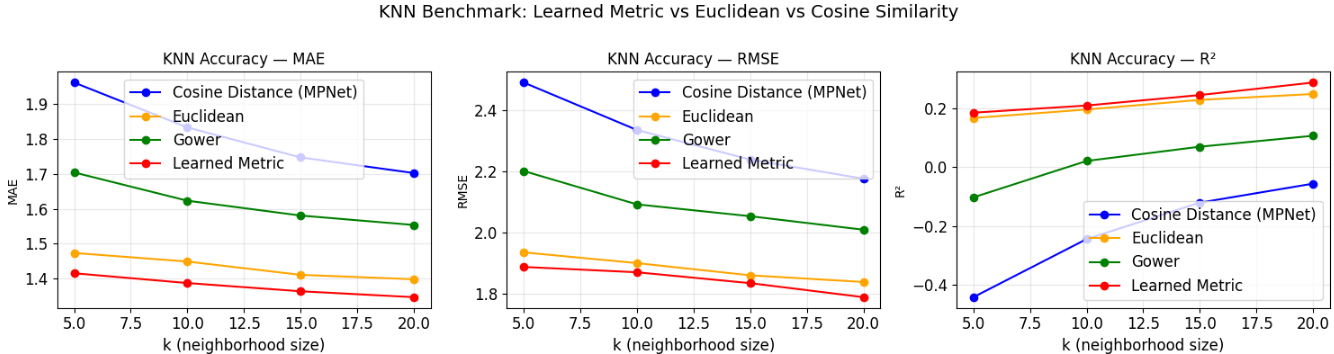}
    \caption{k-NN Valuation by Neighborhood Size and Distance Metric}
    \label{fig:knn_benchmark}
\end{figure*}

In Financial Services (Figure~\ref{fig:knn_benchmark}), the MAE and RMSE of the post-money valuation resulting from the learned metric are consistently lower than the other metrics across all neighborhood sizes. 

\section{Conclusion and Outlook}

We introduce a valuation-conditioned similarity framework for private markets that learns peer relationships from observed transaction valuations. The method combines CatBoost-based valuation modeling with importance-weighted leaf-node co-occurrences to construct an interpretable similarity structure. In the evaluated industry groups, the learned measure produces economically coherent peer sets and generally outperforms conventional distance-based and embedding-based baselines in downstream valuation tasks.

The framework offers a data-driven alternative to comparable-company analysis based on fixed industry, geographic, or deal-stage filters. Rather than matching firms by surface characteristics, it identifies peers that share common valuation drivers.

Future work may extend the approach to scalable cross-industry similarity, softer tree-based proximity measures, and richer textual or graph-based representations of private companies. These extensions could further improve peer identification while preserving the valuation-centered objective of the framework.

\section{Acknowledgement}
The views expressed here are those of the authors alone and not of BlackRock, Inc.

\bibliography{private_market_similarity}

@inproceedings{vamvourellis2024company,
  title={Company similarity using large language models},
  author={Vamvourellis, Dimitrios and T{\'o}th, M{\'a}t{\'e} and Bhagat, Snigdha and Desai, Dhruv and Mehta, Dhagash and Pasquali, Stefano},
  booktitle={2024 IEEE Symposium on Computational Intelligence for Financial Engineering and Economics (CIFEr)},
  pages={1--9},
  year={2024},
  organization={IEEE}
}

@article{rhodes2023geometry,
  title={Geometry-and accuracy-preserving random forest proximities},
  author={Rhodes, Jake S and Cutler, Adele and Moon, Kevin R},
  journal={IEEE Transactions on Pattern Analysis and Machine Intelligence},
  volume={45},
  number={9},
  pages={10947--10959},
  year={2023},
  publisher={IEEE}
}

@article{geertsema2023relative,
  author  = {Geertsema, Paul and Lu, Helen},
  title   = {Relative Valuation with Machine Learning},
  journal = {Journal of Accounting Research},
  year    = {2023},
  volume  = {61},
  number  = {1},
  pages   = {329--376},
  doi     = {10.1111/1475-679X.12464}
}

@article{zhang2023application,
  author  = {Zhang, Ruling and Tian, Zengrui and McCarthy, Killian J. and Wang, Xiao and Zhang, Kun},
  title   = {Application of machine learning techniques to predict entrepreneurial firm valuation},
  journal = {Journal of Forecasting},
  year    = {2023},
  volume  = {42},
  number  = {2},
  pages   = {402--417},
  doi     = {10.1002/for.2913}
}

@article{mehta2025clustering,
  title={Clustering and Similarity Learning in Financial Markets: A Tutorial for the Practitioners},
  author={Mehta, Dhagash and Thompson, John RJ and Lee, Hoyoung and Lee, Yongjae},
  journal={Available at SSRN 5587353},
  year={2025}
}

@inproceedings{cao2024companykg,
  title={Companykg: A large-scale heterogeneous graph for company similarity quantification},
  author={Cao, Lele and von Ehrenheim, Vilhelm and Granroth-Wilding, Mark and Anselmo Stahl, Richard and McCornack, Andrew and Catovic, Armin and Cavalcanti Rocha, Dhiana Deva},
  booktitle={Proceedings of the 30th ACM SIGKDD Conference on Knowledge Discovery and Data Mining},
  pages={4816--4827},
  year={2024}
}

@article{bellet2014survey,
  title={A Survey on Metric Learning for Feature Vectors and Structured Data},
  author={Bellet, Aur{\'e}lien and Habrard, Amaury and Sebban, Marc},
  journal={arXiv preprint arXiv:1306.6709},
  year={2014}
}

@book{damodaran2011damodaran,
  title={Damodaran on valuation: security analysis for investment and corporate finance},
  author={Damodaran, Aswath},
  year={2011},
  publisher={John Wiley \& Sons}
}

@article{shafer2008tutorial,
  title={A tutorial on conformal prediction.},
  author={Shafer, Glenn and Vovk, Vladimir},
  journal={Journal of machine learning research},
  volume={9},
  number={3},
  year={2008}
}

@article{Breiman2001Random,
  title={Random Forests},
  author={Breiman, Leo},
  journal={Machine Learning},
  volume={45},
  number={1},
  pages={5--32},
  year={2001},
  publisher={Springer}
}

@article{Friedman2001Greedy,
  title={Greedy Function Approximation: A Gradient Boosting Machine},
  author={Friedman, Jerome H.},
  journal={Annals of Statistics},
  volume={29},
  number={5},
  pages={1189--1232},
  year={2001}
}

@article{Lundberg2020From,
  title={From Local Explanations to Global Understanding with Explainable AI for Trees},
  author={Lundberg, Scott M. and Erion, Gabriel G. and Chen, Hugh and others},
  journal={Nature Machine Intelligence},
  volume={2},
  pages={56--67},
  year={2020}
}

@inproceedings{xing2002distance,
  title={Distance Metric Learning with Application to Clustering with Side-Information},
  author={Xing, Eric W and Jordan, Michael I and Russell, Stuart J and Ng, Andrew Y},
  booktitle={Advances in Neural Information Processing Systems (NeurIPS)},
  volume={15},
  pages={521--528},
  year={2002}
}

@article{weinberger2009distance,
  title={Distance Metric Learning for Large Margin Nearest Neighbor Classification},
  author={Weinberger, Kilian Q and Saul, Lawrence K},
  journal={Journal of Machine Learning Research (JMLR)},
  volume={10},
  number={2},
  pages={207--244},
  year={2009}
}

@inproceedings{Saha2024,
  title={Machine learning-based relative valuation of municipal bonds},
  author={Saha, Preetha and Lyu, Jasmine and Desai, Dhruv and Chauhan, Rishab and Jeyapaulraj, Jerinsh and Chu, Peter and Sommer, Philip and Mehta, Dhagash},
  booktitle={Proceedings of the 5th ACM International Conference on AI in Finance},
  pages={634--642},
  year={2024}
}

@inproceedings{Jeyapaulraj2022,
  author    = {Jeyapaulraj, Jerinsh and Desai, Dhruv and Chu, Peter and Mehta, Dhagash and Pasquali, Stefano and Sommer, Philip},
  title     = {Supervised Similarity Learning for Corporate Bonds Using Random Forest Proximities},
  booktitle = {Proceedings of the 3rd ACM International Conference on AI in Finance (ICAIF '22)},
  pages     = {411--419},
  year      = {2022},
  publisher = {ACM}
}

@article{rosaler2025supervised,
  title={Supervised similarity for high-yield corporate bonds with quantum cognition machine learning},
  author={Rosaler, Joshua and Candelori, Luca and Kirakosyan, Vahagn and Musaelian, Kharen and Samson, Ryan and Wells, Martin T and Mehta, Dhagash and Pasquali, Stefano},
  journal={arXiv preprint arXiv:2502.01495},
  year={2025}
}

@misc{edlich2026privateequity,
  author       = {Edlich, Alexander and Llewellyn, Chris and Croke, Christopher and Schneider, Rahel and Teichner, Warren},
  title        = {Private Equity --- Global Private Markets Report},
  institution  = {McKinsey \& Company},
  year         = {2026},
  month        = {February},
  url          = {https://www.mckinsey.com/industries/private-capital/our-insights/global-private-markets-report/private-equity},
  note         = {Accessed: May 2026}
}

@incollection{lundberg2017shap,
title = {A Unified Approach to Interpreting Model Predictions},
author = {Lundberg, Scott M and Lee, Su-In},
booktitle = {Advances in Neural Information Processing Systems 30},
editor = {I. Guyon and U. V. Luxburg and S. Bengio and H. Wallach and R. Fergus and S. Vishwanathan and R. Garnett},
pages = {4765--4774},
year = {2017},
publisher = {Curran Associates, Inc.},
url = {http://papers.nips.cc/paper/7062-a-unified-approach-to-interpreting-model-predictions.pdf}
}

@article{yang2021fast,
  title={Fast TreeSHAP: Accelerating SHAP Value Computation for Trees},
  author={Yang, Jilei},
  journal={arXiv preprint arXiv:2109.09847},
  year={2021}
}

@inproceedings{akiba2019optuna,
  title={Optuna: A next-generation hyperparameter optimization framework},
  author={Akiba, Takuya and Sano, Shotaro and Yanase, Toshihiko and Ohta, Takeru and Koyama, Masanori},
  booktitle={Proceedings of the 25th ACM SIGKDD international conference on knowledge discovery \& data mining},
  pages={2623--2631},
  year={2019}
}

@inproceedings{jeyapaulraj2022supervised,
  title={Supervised similarity learning for corporate bonds using Random Forest proximities},
  author={Jeyapaulraj, Jerinsh and Desai, Dhruv and Mehta, Dhagash and Chu, Peter and Pasquali, Stefano and Sommer, Philip},
  booktitle={Proceedings of the Third ACM International Conference on AI in Finance},
  pages={411--419},
  year={2022}
}

@article{prokhorenkova2018catboost,
  title={CatBoost: unbiased boosting with categorical features},
  author={Prokhorenkova, Liudmila and Gusev, Gleb and Vorobev, Aleksandr and Dorogush, Anna Veronika and Gulin, Andrey},
  journal={Advances in neural information processing systems},
  volume={31},
  year={2018}
}

@inproceedings{li2025similarity,
  title={Similarity-based Conformal Prediciton using Random Forest Proximities},
  author={Li, Mingshu and Desai, Dhruv and Sarmah, Bhaskarjit and Bhagat, Snigdha and Mehta, Dhagash},
  booktitle={Proceedings of the 6th ACM International Conference on AI in Finance},
  pages={387--395},
  year={2025}
}
\bibliographystyle{unsrtnat}
\appendix


\end{document}